\documentclass[twocolumn]{article}
\usepackage{graphicx} 
\usepackage[T1]{fontenc}
\usepackage{hyperref}
\usepackage{algorithmicx}
\usepackage{babel}
\usepackage{array}
\usepackage[backend = bibtex, style = numeric, maxbibnames=99]{biblatex}
\title{Prompting Code Interpreter to Write Better Unit Tests on Quixbugs Functions}

\author{
  Vincent Li\footnote{University of Chicago Existential Risk Laboratory (institution where work was primarily conducted)}  \footnote{ Yale University (author’s current institution)}\\
  \texttt{mail2vincentrli@gmail.com}
  \and
  Nick Doiron\footnote{Hewlett Packard Enterprise}\\
  \texttt{nick.doiron@hpe.com}
}

\date{}

\begin{document}

\maketitle

\begin{abstract}
    Unit testing is a commonly-used approach in software engineering to test the correctness and robustness of written code. Unit tests are tests designed to test small components of a codebase in isolation, such as an individual function or method. Although unit tests have historically been written by human programmers, recent advancements in AI, particularly LLMs, have shown corresponding advances in automatic unit test generation. In this study, we explore the effect of different prompts on the quality of unit tests generated by Code Interpreter, a GPT-4-based LLM, on Python functions provided by the Quixbugs dataset, and we focus on prompting due to the ease with which users can make use of our findings and observations. We find that the quality of the generated unit tests is not sensitive to changes in minor details in the prompts provided. However, we observe that Code Interpreter is often able to effectively identify and correct mistakes in code that it writes, suggesting that providing it runnable code to check the correctness of its outputs would be beneficial, even though we find that it is already often able to generate correctly-formatted unit tests. Our findings suggest that, when prompting models similar to Code Interpreter, it is important to include the basic information necessary to generate unit tests, but minor details are not as important. \newline \newline \newline
\end{abstract}

\section{Introduction}
\par In software engineering, testing the correctness of written code, especially before deployment, is of utmost importance, since it greatly reduces the possibility of unexpected errors and crashes. A common approach to software testing is \textit{unit testing}, in which code is broken down into smaller components whose correctness can be tested individually. Often, this is done by individually testing a \textit{focal method} or a \textit{focal class} in isolation.
The advantage of such an approach is that breaking down code into smaller components reduces its complexity, making it easier for human programmers to construct a comprehensive unit test suite that includes a diverse set of edge cases. Furthermore, it allows human programmers to more easily pinpoint the location and cause of errors and discrepancies between the expected and actual output of the code, thus facilitating the debugging process. However, writing unit tests is often a time-consuming process that therefore demands a large portion of a developer’s time and energy. 
\par In recent years, with the rise of Large Language Models (LLMs) such as ChatGPT \cite{openai_introducing_2023}, there has been an increasing focus on the application of LLMs to the task of writing unit tests, as they have the potential to drastically reduce the time necessary to properly and sufficiently test written code before deployment. Therefore, the study of the unit test generation capabilities of LLMs has the potential to provide a valuable tool for developers and to greatly increase the speed at which developers can produce correct code that handles edge cases well.

\subsection{Related Work}

\par In recent years, several previous works have attempted to tackle the problem of automatic unit test generation, with many opting to use LLMs for this purpose. For instance, AthenaTest \cite{tufano_unit_2020} and A3Test \cite{alagarsamy_a3test_2023} both use transformer-based language models to generate unit tests. Many previous methods, such as TestPilot (based on Codex) \cite{schafer_adaptive_2023}, ChatUniTest \cite{xie_chatunitest_2023}, and ChatTester \cite{yuan_no_2023}, (both based on ChatGPT) make use of an iterative algorithm, in which a code generation model is initially prompted to generate unit testing code, and if there are errors in the generated code, it is repeatedly prompted with the goal of inducing it to fix the errors. Non-iterative methods include differential prompting \cite{li_finding_2023}, in which the model is prompted to find failing test cases for a given focal method by generating multiple reference implementations and finding the test cases in which the method under test and the reference implementations produce different results.

\par \textbf{Common Failure Modes of Generated Unit Tests.} Although unit tests generated by LLMs have the advantage of being less time-consuming to generate, as compared to those generated by humans, they have the disadvantage of being more likely to contain errors that prevent successful compilation of the code. Indeed, such syntax and compilation errors are a significant source of ineffective unit testing code \cite{schafer_adaptive_2023, xie_chatunitest_2023, yuan_no_2023}. Other sources of ineffective unit tests include the unit testing code running for longer than the enforced timeout or asserting incorrect values, the latter of which may cause correct focal functions to be marked as incorrect and incorrect focal functions to be marked as correct \cite{schafer_adaptive_2023}. However, due to the prevalence of compilation errors and the necessity of compilable code, it is imperative that a reliable method of fixing or correcting these errors be found.

\par \textbf{Prompting.} There are several components of the prompts given to LLMs to generate unit tests that have been investigated by previous works.
\par Due to the prevalence of iterative methods in the literature \cite{schafer_adaptive_2023, xie_chatunitest_2023, yuan_no_2023}, the investigation of these methods sheds much light on the variety of different variables, enumerated below, that may be modified to alter the quality of the generated unit tests. Iterative methods work by first generating an initial prompt, and then subsequently reprompting the model to correct errors or otherwise achieve a better result. Even though not all methods are iterative, we will, to the ends of readability and ease of characterization, simply treat them as iterative methods that only generate an initial prompt, without any reprompting. 

\begin{enumerate}
    \item \textit{Amount of Code Context Given in the Initial Prompt.} It is important that the user provide the model with enough information (\textit{i.e.} code context, such as the definition of the focal function or focal method, the signature of the focal function, etc.) to write proper unit tests, and in general, it is better to give more information, while keeping in mind the limited size of the model’s context window \cite{schafer_adaptive_2023, tufano_unit_2020}. In particular, it is common practice to include information about the focal method’s body in the initial prompt, whether that information be included directly, in the prompt itself \cite{tufano_unit_2020, xie_chatunitest_2023}, or indirectly, such as by using that information to write an NL (natural language) description of the focal method’s intention, which is then used as the actual prompt \cite{yuan_no_2023, li_finding_2023}. It follows from the evidence in previous studies that this leads to better results. 
    \par Aside from the information about the focal method’s signature and body, other relevant pieces of information include the \textit{focal class} (the class that the focal method is a part of, if any), the fields and methods of the focal class, and the \textit{dependencies} of the focal method (the functions called by the focal method). Here, there is less consensus among previous studies. Some include that information in the initial prompt directly, whether it be with dependencies \cite{xie_chatunitest_2023} or without \cite{tufano_unit_2020}; some include it indirectly (via the generation of an intermediate NL prompt describing the intention) \cite{yuan_no_2023}; and others do not include it at all, in the initial prompt (although they may or may not include it in subsequent prompts) \cite{li_finding_2023, schafer_adaptive_2023}. 

    \item \textit{Code Documentation.} TestPilot \cite{schafer_adaptive_2023} also draws upon the documentation of the focal methods used. In particular, it, at times, prompts the model with doc comments (NL comments in the documentation that describe the focal method and its intention) and examples of the focal method’s usage. However, the drawback of this approach is that it is only possible when there exists documentation from which to draw this information. For many focal methods, the relevant documentation does not exist. 
    \par However, we may instead inquire about the effectiveness of including documentation information of the dependencies, especially if the dependencies include functions from well-known and widely-used libraries. Perhaps this may be an effective strategy. Alternatively, if a focal method or one or more of its dependencies does not have human-written documentation available online, it may possibly be beneficial to prompt the model (or a different model) to write documentation-style comments and usage examples for those functions itself, and then prompt the model with the output. Even though past studies have explored the possibility of having the model write an NL comment describing the focal method’s purpose \cite{yuan_no_2023}, the question of how the other prompting methods would affect the quality of the generated unit tests remains, though it is beyond the scope of this study.
    \item \textit{Followup Prompting.} In iterative methods, it is important to design the subsequent prompts to the model to maximize their effectiveness. Due to the prevalence of compilation errors in the generated unit test code, the primary objective of reprompting is to lead the model to repair errors in its code. 
    \par In the literature, we have found three distinct factors that previous studies have experimented with: (1) the amount of information, or additional information, about the code context to include, (2) whether error messages are directly fed back into the model, or whether they are converted to feedback given in NL, and (3) the model involved in reprompting, if any. 
    \par With respect to information included in subsequent prompts, previous models have either included only information related to the error message \cite{yuan_no_2023, xie_chatunitest_2023}, even for general coding applications outside of unit testing \cite{olausson_demystifying_2023}, or opted to include additional code context information due to finding it to be a more effective strategy \cite{schafer_adaptive_2023}.
    \par Of the studies whose proposed reprompting algorithms include only the information about the error messages, some directly include the error message as part of the prompt \cite{xie_chatunitest_2023, yuan_no_2023}. Other studies experiment with using LLMs or human feedback to turn the error message into NL before reprompting the code-generating model, albeit on general coding tasks \cite{olausson_demystifying_2023}. 
    \par If a subsequent prompt is constructed by taking the error message, as is, and taking that as part of the subsequent prompt, then it does not invoke an LLM to generate that prompt. However, evidence suggests that using more advanced models, such as GPT-4, to turn the error message into an NL explanation of the error may be more effective \cite{olausson_demystifying_2023}. 

\end{enumerate}

\par \textbf{Prompt Factors.} Through a combination of reviewing the literature and independently brainstorming, we come up with and present a list of factors that may cause prompts to result in different outputs in Appendix A. It is our hope that future researchers may find this list useful during literature review; however, we include it in the appendix for the sake of readability. \newline
\par \textbf{The Present Study.} Although there exist a variety of different factors, such as model size, training data quality, and code quality, that affect the quality of generated unit tests, the present study focuses on the effect of prompting on the quality of the generated tests. There are several reasons for this. Firstly, the prompt is one of the elements of the model over which the user is able to exercise a great deal of control. Therefore, our findings may be applied by any user, rather than only those with access to the internals of the model and with the ability to change them. Secondly, the ease at which users can change prompts means that they are able to tailor the prompts to suit their own unique needs and preferences. Thirdly, a study of the kinds of prompts that are effective in eliciting high-quality unit tests may provide useful insight into the nature of the model, since it may elucidate helpful general principles for eliciting high-quality responses. Such studies are also relatively uncommon in the literature.
\par In the present study, we study the ability of Code Interpreter \cite{openai_code_2023}\footnote{Note that the usage is capped like with the default version of GPT-4 (the message displayed is that the usage of GPT-4 has reached the limit), and that the user interface (https://chat.openai.com/?model=gpt-4-code-interpreter, as of August 2, 2023) classifies it under GPT-4, despite the blog post calling the underlying model a ChatGPT model, so we assume it to be a GPT-4 model.
}, a GPT-4 model with the ability to generate code, to generate unit tests for functions in the Quixbugs dataset \cite{lin_quixbugs_2017}. In particular, we focus on how to engineer the prompt to achieve optimal results. We find evidence to suggest that, when prompt-engineering with Code Interpreter, the quality of the model’s outputs, according to multiple metrics, is not sensitive to changes in minor, less relevant details in the prompt. Therefore, our findings suggest that users of Code Interpreter need not consider the details, so long as the basic information necessary for the model to perform its task are contained in the prompt.

\section{Methodology}
\par The purpose of this study is to investigate how using different prompts affects the quality of generated unit tests. Therefore, we first generate the contents of the prompt that is fed into Code Interpreter, and then evaluate the output. Rather than giving subsequent prompts after the initial prompt, we only give it an initial prompt and evaluate the output from that prompt. This noniterative workflow of giving one prompt and receiving one reply per conversation simplifies the evaluation process and allows us greater freedom in experimenting with having the model regenerate its response to the same prompt multiple times, thus giving a more balanced and comprehensive view of its response to any given prompt.

\subsection{Prompt Generation}
\par When generating the content of the prompts, we consider the following dimensions along which the prompts may vary (for more details, see Appendices B and C):

\begin{enumerate}
    \item \textit{Format of Code Context.} We experiment by giving it the code context in either one of two formats: NL format, as a description of the function, or as the code of the focal function itself. The NL description is generated by Code Interpreter itself from the code of the incorrect implementation of the function body and simply copied and pasted into the prompt for generating the unit tests. In order to prevent the model from using that code when generating unit tests, we ask it to generate unit tests in a separate conversation window.
    \par The code itself comes from the Quixbugs dataset \cite{lin_quixbugs_2017}, which gives both a correct and incorrect implementation of each focal function. Because unit tests are conducted for the purpose of finding implementation errors, we give the incorrect implementation of the function when we give context in code format. All code is given in Python. 
    \par Regardless of the format of the code context, we always include the function signature, which includes the function name and the names of the inputs.
    \item \textit{Number of Example Unit Tests Provided.} 	We experiment with giving the model formatting examples of zero, one, or two separate unit test cases. Regardless of the number of unit test case examples provided, we always provide adequate instruction for the expected format of the unit test cases, and in the prompt, we encourage the model to generate unit tests that are comprehensive and cover edge cases.
    \item \textit{Different Focal Functions.} Although it would be ideal to prompt with all of the focal functions in the dataset, we removed some for various reasons, such as: to ensure that the remaining functions were amenable to testing solely their inputs and outputs (\textit{i.e.} they did not require testing of the intermediate processes, as was the case in functions such as depth- and breadth-first search, where the order of search is relevant), to filter out those for which a given input could have multiple correct outputs, to exclude those for which we suspected that the given correct implementation may not have been correct, to avoid functions whose inputs or outputs contained formatting not compatible with the testing framework producing accurate results (such as difficult-to-parse tuple or string expressions), etc. The subset of functions that remains can be found in Appendix D.
    \item \textit{Miscellaneous NL Comments.} We test whether the model produces better-written unit tests and catches the mistake in the incorrect implementation more often if the prompt contains comments such as “You are an expert programmer”, which thus creates two distinct possibilities for the kinds of NL comments in the prompt. 
\end{enumerate}
\par To see the code that was used to generate the prompts, and to view the prompt-generation process in more detail, please refer to Appendices B and C. We probe the model using every combination of the above 4 dimensions to generate prompts. For each prompt, we sample the model’s output a total of 5 times. In cases when the model’s output length exceeds the length of the preset token limit for a single response, we simply allow it to continue generating until it finishes its response. We collect data from August 1–16, 2023, inclusive.

\subsection{Output Evaluation}
\par When evaluating the output produced by the model, we check several components of the output:

\begin{enumerate}
    \item \textit{Correctness of Format.} We check whether the format, whether given as a downloadable file or an embedded code block, conforms to the format specified in the prompt. If it does not, then we do not use those test cases, as they are incompatible with the testing framework. Therefore, all data about the provided test cases comes solely from those generated with the correct format.
    \item \textit{Whether the Mistake is Corrected.} We observe that the model will sometimes catch the mistake in the incorrect implementation. We consider the model to have corrected the mistake if and only if it either correctly re-implements the function or it points out the error and the exact changes necessary to fix it.
    \item \textit{Correctness of Test Cases.} We check whether the expected output that the model gives in the test cases matches the output of the correct implementation of the function. If so, then we consider the test case to be correct. 
    \item \textit{Coverage of Test Cases.} We examine whether the given test cases are \textit{failing}, which we define as a test case in which the correct and incorrect implementations give different outputs. Therefore, failing test cases are those that are capable of uncovering errors in the incorrect implementation, if the correct output is known. Note that a failing test case is failing, regardless of whether the expected output, given in the test case, is correct; the input is the important variable for determining whether a test case is failing.
    \item \textit{Correct Coverage.} Because it is optimal for failing test cases to also be correct, we also combine the above two points by checking which test cases are both correct and failing.
\end{enumerate}

\par \textit{Error Handling} When the function throws an error as a result of the input from a test case, we handle it as follows: if the correct implementation throws an error, regardless of the behavior of the incorrect implementation, then we assume that the input does not satisfy the function’s precondition, which means the function’s behavior is undefined. Therefore, we consider the test case to be correct, since any output is correct, and not failing, because causing undefined behavior does not reveal the incorrectness of the incorrect implementation. 
\par If the correct implementation does not throw an error, but the incorrect implementation throws, then we consider the test case to be a failing test case. 

\section{Results}
\par We present data about all of the test cases, as whole, in Table 1. Based on the data, in order to have at least 10 expected correct failing test cases, it is advisable to resample at least 4 times from the same prompt. 

\begin{center}
\begin{table}[]
    \centering
    \begin{tabular}{ m{1.4cm} || b{1.4cm}  b{1.4cm}  b{1.4cm}  b{1.4cm}}
        \hline
        & Mean No. of TCs per Response & Standard Deviation in No. of TCs per Response & Mean Frac. of TCs & St. Dev. in Frac. of TCs  \\
        \hline 
        & & & & \\
        CFORM & 7.78 & 3.95 & 1.00 & 0.00 \\
        & & & & \\
        CORR & 6.13 & 4.18 & 0.79 & 0.54 \\
        & & & & \\
        FAIL & 4.43 & 3.72 & 0.57 & 0.48 \\
        & & & & \\
        CORR and FAIL & 3.11 & 3.59 & 0.40 & 0.46 \\
        \hline 
    \end{tabular}
    \caption{Data collected about the test cases together, as a whole. The term fraction refers to the number of test cases with the given characteristic, per correctly-formatted test case. For example, the mean fraction of test cases per response for correct test cases refers to the proportion of test cases that were correct (on average, considering only correctly-formatted responses), which is computed by first aggregating the data and then averaging, rather than in the opposite order. Abbreviations used: CORR: correct test cases; FAIL: failing test cases; CFORM: correctly formatted; No.: number; frac.: fraction; st. dev.: standard deviation; TC: test case.}
    \label{tab:my_label}
\end{table}
\end{center}

\par Upon performing t-tests on the resultant data, we find that, with a few exceptions, between any two types of prompt, the variability from reprompting and the diversity of focal functions provided was substantially greater than that provided by differences in prompting style. In other words, for most prompts, there was no significant evidence of a difference in any of the measured quantities arising from changes to the prompting style\footnote{ The exceptions are that prompts that included the code, did not include miscellaneous NL comments, and included 1 or 2 output examples showed significantly more correctly formatted outputs than prompts that include an NL function description, miscellaneous NL comments, and no output examples (p = 0.000039 if the former prompt has 1 example; p = 0.000042 if the former prompt has 2 examples, which is significant at the alpha = 0.05 level). However, despite the low p-values of these differences, we do not focus much attention on their significance because the mean difference is small, and because we do not believe that this will bring much of a change to the end user, due to it being a difference in only a single metric.}.
\par However, despite this, in general, we find that prompts that give code context directly (\textit{i.e.} as code), do not include miscellaneous NL comments, and include 2 output examples are more generally associated with better performance metrics, while prompts that include the code context as an NL description, include miscellaneous NL comments, and do not have output examples have the opposite effect, though we note that the difference is not large.
\par Thus, we advise that future Code Interpreter users employ the former prompting style when creating prompts. 
\par For the complete data collected, please see Appendix E.

\subsection{Observations}
\par In addition to the data above, we also make several qualitative observations about the responses produced by Code Interpreter, especially in regards to the format of the responses generated.
\par Firstly, we observe that, when the model generates test cases that are not in the format that we specified, it is often in a similar, but more standard, format. In fact, the model response will sometimes point out that the format we specify is nonstandard JSON. Therefore, it appears that the model has an inclination towards working with more standard formats, suggesting that it would be beneficial to ask for standard, rather than nonstandard, formats, unless there is strong reason to use a nonstandard format, such as testing the model’s ability to generalize to nonstandard formats that do not appear often in the training data.
\par Secondly, the model is often not able to format test cases with multiple or nested sets of brackets correctly. Often, this happens in the test cases for sorting functions, since their input is a list of numbers (\textit{i.e.} [1, 2, 3]), rather than multiple numbers (\textit{i.e.} 1, 2, 3). More generally, the model often produces test cases that are marked as incorrect by the testing framework, but would have been correct if they were correctly formatted. Thus, it sometimes misunderstands the expected test case format, despite the formatting instructions that we provide.
\par Thirdly, and most interestingly, when the model itself writes Python code containing an error, it will often be able to correctly explain the error message and make the corresponding correction because it has access to its own internal Python environment. Therefore, to increase the probability of the model producing correctly-formatted test cases, we conjecture that it would be helpful to provide Python code designed to check whether the test cases are correctly formatted, allowing the model to check and correct the format of the cases that it generates. Even though providing explicit and specific formatting instructions helps the model produce correctly-formatted test cases, we conjecture that providing the format-checking code will be more effective, especially considering that the model does not always follow the formatting instructions. Alternatively, asking the model to provide the unit test as Python code may also be effective, since it would be able to run that code in its integrated Python environment. However, when the focal function causes an assertion error in the unit test code (which could happen when the actual output does not match the expected output) it would be important for the model to be able to differentiate that kind of error from an error in the unit test code itself.

\section{Discussion}
\par Although previous works with previous LLMs have found that the quality of those models’ outputs are noticeably influenced by the modification of details in the prompts, our findings are to the contrary, suggesting that the present model under test is more robust to such changes. Even though we do not test for all of the factors suggested in previous studies, we posit that Code Interpreter will not be as sensitive to such details as previous models are. 
\par However, it is important to keep in mind that there exist several basic elements that have always been present in all of our prompts. For example, we give clear instructions for how we expect the output to be formatted, a description of the focal function derived from its code (if not the code of the focal function itself), and resampled multiple outputs from the same initial prompt. We believe these elements to be necessary for the model to produce an output that conforms to the user’s requirements.
\par We speculate that this change from previous models is due to a difference in the training process that GPT-4 underwent, as compared to previous models, which thus allowed it to become robust to less relevant details and only be affected by more important details, much like a human programmer would. Because unit test generation is a well-defined and relatively common task for programmers, the abundance of available training data would only have served to accelerate this effect. However, without knowing more about the training process, it is difficult to make a definitive statement.
\par As AI nears or surpasses human levels of intelligence in increasingly many domains, it is ever more important to pay close attention to the risks inherent in AI. 

\subsection{Limitations}
\par As with any study, there exist limitations to ours, which we describe and address below.
\par First, we focus primarily on resampling outputs from the same initial prompt, rather than using followup prompts to correct or enhance the existing output. However, we note that our approach does not require that we tailor a followup prompt specifically suited to induce the model to correct the existing output. Furthermore, the need to generate a followup response in order to repair the original response is also fulfilled by regenerating more responses from the initial prompt. This is because, of the multiple responses generated, it is exceedingly likely that at least one is of the correct format, since generating more responses will increase the chances that at least one is in the correct format.
\par Second, our testing framework is limited by the kinds of unit tests that it can run. Therefore, we picked a subset of the functions that were compatible with the testing framework. This subset was characterized by functions that were amenable to simple input/output testing. However, a study on how the results might change if a more diverse set of functions (such as those requiring an in-depth examination of the intermediate stages, like breadth- and depth-first search) and test cases (such as those in which one of multiple possible answers could be correct) were included would be quite informative. Furthermore, an improvement in the way that our testing framework handled errors thrown by the focal function (either the correct or incorrect implementation) would also have provided a more accurate view of the behavior of both implementations under the generated test cases.

\subsection{Future Work}
\par An interesting future research direction to explore would be to investigate how followup prompts affect the quality of generated unit tests, and how such prompts can be designed to most efficiently do so. Furthermore, it would be worth investigating how models fare against more complex codebases containing multiple, interdependent functions or functions that are not commonly in the training data, especially if their doc comments are sparse or nonexistent.
\par Additionally, a study on the degree to which the test cases were memorized from the training data, or how the prompt could be improved if the model were asked to improve it, would be informative.

\subsection{Alignment and Risks}
\par With the rise of AI capabilities, AI developers and researchers must take care to prevent the misuse and misalignment of increasingly powerful AI systems, especially as they near the level of artificial general intelligence (AGI). If AGI is realized, then misaligned agents have the potential to cause great, possibly even existential, harm to society, especially considering the increasing reliance on AI systems. 
\par Our work seeks to address this by shedding light on best practices for prompting AI models to construct more rigorous and comprehensive unit tests. Although there is still much room for improvement before certifiably safe and aligned AGI systems can be made, it is our hope that our work can make and inspire progress towards reaching that goal by laying the foundations for the automatic generation of safety checks through disseminating widely-applicable knowledge of unit test prompting best practices. 
\par However, it is also important to be mindful of unexpected capabilities that emerge from large models and to be prepared for the unforeseen challenges that may arise when attempting to make AI systems aligned. To deal with these challenges as they arise will be crucial for successfully working towards the goal of creating safe and aligned AGI systems.

\section{Conclusion}
\par In this study, we investigate the effect of prompting on the quality of unit tests generated by Code Interpreter. In particular, we vary the format in which the code context was provided, the number of formatting examples for the expected output, and whether we include NL miscellaneous comments in the prompt telling the model that it is an expert coder. Although these factors do not have a significant effect on the quality of the generated unit tests, we make several observations about the outputs of Code Interpreter, the most interesting of which is that its ability to run and correct its own Python code is quite effective and suggests that including code to check for the correctness of the output format would be useful. Future work could explore the effect of followup prompts on the quality of generated unit tests. However, as AI continues to advance, it is important that researchers increasingly focus their attention on preventing harms from AI-related issues, such as misalignment. 

\section{Acknowledgements}
\par We would like to express gratitude to Zachary Rudolph and the University of Chicago Existential Risk Laboratory (XLab) for providing the funding and facilities necessary to conduct this research. In particular, Zack’s mentorship, feedback, and support proved invaluable for this project, and VL feels that the XLab Summer Research Fellowship has imparted on him a substantial amount of knowledge about existential risk, which is an important concern in today’s times, even if he also thinks that Zack calls on him too often.

\printbibliography

\newpage
\section{Appendices}

\subsection{Appendix A: Prompting Factors}
\par Through a combination of reviewing the literature and independently brainstorming, we come up with and present a list of factors that may cause prompts to result in different outputs below. It is our hope that future researchers may find this list useful in their literature review processes.
\par Note that, for some of the factors listed below, more information is given in the Prompting section in the introduction.

\begin{enumerate}
    \item \textit{Amount of Code Context Given.} Inspired by previous work \cite{schafer_adaptive_2023, tufano_unit_2020}, we ask: what is the ideal amount of context to give to the models? Would it be best to include only the minimal amount of information possible, which is only the focal method’s name and signature, or would it be beneficial to also include the focal method’s body, the focal class, its fields and methods, and the dependencies of the focal class? 
    \item \textit{Format of the Code Context.} To build on previous work \cite{yuan_no_2023, li_finding_2023}, we ask: if more code context than the bare minimum is given, what is the best format to give it in? Would it be more beneficial to give it as pure code, or should we prompt the model to generate an NL description of the code context?
    \item \textit{Amount of Outside Documentation Given.}We will assume that outside documentation for the focal method itself does not exist. However, if well-documented library functions exist among the dependencies, should their documentation be included in the prompt? If so, would it be beneficial to prompt the model to summarize the documentation comment, or should it be left in its original form? Do usage examples benefit the model? This line of questioning was inspired by the work done in a previous study in the literature \cite{schafer_adaptive_2023}.
    \item \textit{Number of Example Unit Tests Provided.} In the initial prompt, how many examples of well-written human-made unit tests is optimal? Does the amount of unit tests depend on the amount of code context already given, relative to the size of the model’s context window?
    \item \textit{Information Given in Subsequent Prompting.} Inspired by previous work \cite{schafer_adaptive_2023}, we ask: when followup prompting, is it better to gradually include more code context information in the subsequent prompts, or should the only information given in the subsequent prompts be related to the error message from compilation, if any? 
    \item \textit{Method of Generating Feedback Prompts.} As a follow-up on previous work \cite{olausson_demystifying_2023}, we ask: once the information needed for feedback is given, how should the feedback prompts be generated? Should the error message be fed directly back into the model, or should an NL description of the error message and how to fix it be generated? If so, what model should be used to generate the NL? Is it better to use larger, more capable models? 
    \item \textit{Coding Language Used.} How does the coding language used affect the quality of the generated unit tests? A previous study in the literature has found that multilingual, NL LLMs often develop internal representations that consist of a single language-agnostic component and multiple language-specific components, for each language that the model uses \cite{choenni_how_2023}. Although the results apply for natural language models, whether they also extend to coding languages would be worth investigating. If so, this type of representation would suggest that, due to the commonality of the language-agnostic component between languages, the language model may be able to adapt to programming languages that do not appear in its training data, since it will still be able to make use of its knowledge of the underlying programming concepts. 
    \par If this is so, then how many code examples in the prompt are sufficient to give the model enough information about the new programming language to be proficient enough to write unit tests? Would it require NL explanations of the code?  
    \item \textit{Resampling vs. Iterative Correction.} This study \cite{olausson_demystifying_2023} suggests that, in some contexts, higher-quality code may be generated by simply giving the model an initial prompt, and taking advantage of the inherent non-determinism in LLMs by eliciting it to generate many responses from the same initial prompt (resampling), rather than by iteratively correcting a few responses. In the context of unit-test generation, would it be beneficial to prioritize resampling or iterative correction?
    \item \textit{Miscellaneous NL Comments.} Does including NL comments in the prompt such as “You are an expert programmer” cause the model to produce more expertly-written responses? Do comments with the opposite meaning induce the opposite response?
\end{enumerate}

\subsection{Appendix B: Prompt-Generating Code}
\par We share the link to a Google Colab IPython notebook below and divide the code into sections for ease of understanding. As a whole, the notebook code serves multiple purposes, including generating the model prompts, evaluating the test cases given by the model, processing the data, and saving the data to Google Drive.
\newline 
\url{https://colab.research.google.com/drive/1WthhcG9D\_6Uf8jRjRtYKHLNlM5JuFhxb?usp=sharing}

\subsection{Appendix C: Prompt Template}
\par Below, we present templates of the prompts that we used, indicating information to fill in italicized and with curly braces.

\par \textbf{Pre-Prompt.} We only give the model a pre-prompt if we seek to induce it to first write an NL description of the code before prompting it to write the unit tests. The pre-prompt template is below:
\newline

\begin{ttfamily}
    Please describe the intent and purpose of the following Python function in natural language: \textit{\{function signature\}}. The code of the function is below: \\ \\

<BEGIN CODE> \\
\textit{\{code of the incorrect implementation of the function\}}\\ \\

<END CODE> \\ \\
\end{ttfamily}

\par \textbf{Initial Prompt.} The initial prompt, which is used to prompt the function to generate unit test cases, is below: \newline 

\begin{ttfamily}
    \textit{\{If including miscellaneous NL comments, then include: “You are a world-class programmer and an expert in Python.”\} }Please generate a compilable and comprehensive suite of unit tests for the following Python function: \textit{\{function signature\}}. The unit tests should be written to cover edge cases. The code of the function is below: \\ \\

<BEGIN CODE> \\
\textit{\{Code of the incorrect implementation of the function. If, instead, an NL description of the function is being used instead of the code, then the NL description goes here, and the section markers “<BEGIN CODE>” and “<END CODE>” are replaced with “<BEGIN DESCRIPTION>” and “<END DESCRIPTION>”, respectively. We do not include both the code and NL description in the same prompt.\}} \\ \\

<END CODE> \\
Please generate the test cases as a JSON file in the following format: `[[input(s)], expected output]'. Each test case should be on its own line. Do not include any comments in the test cases. Do not separate individual test cases with commas. Do not include the outer set of brackets in the JSON test cases. \\ \\

\textit{\{If including 1 example for output formatting, then include: “For example, if the input to a test case is `[1,1,1,134]' and the expected output is `false', then the correct format is `[[[1,1,1,134]], false]'.”\}} \\ \\

\textit{\{If including 2 examples for output formatting, then include the above and the following (note that “\textbackslash n” represents a newline character): “Another example: if the input is `234,1,1' and the expected output is `[971,1,0]', then the correct format is `[[234,1,1], [971,1,0]]'. Together, both test cases should be written as \textbackslash n ``` \textbackslash n[[1,1,1,134], false] \textbackslash n[[[234]], [971,1,0]]\textbackslash n''' ”\}} \\ \\ 

\textit{\{If including miscellaneous NL comments, then include: “Remember, you a seasoned and experienced programmer.”\}} \\ \\

\end{ttfamily}

\subsection{Appendix D: Functions Tested}
\par The 20 functions from Quixbugs that we tested are below. Note that, to get to the file name, simply append “.py” to the function name and search in the folder \texttt{Quixbugs/python\_programs}. \newline \newline 
\begin{ttfamily}
    bucketsort \\
    find\_first\_in\_sorted \\
    find\_in\_sorted \\
    gcd \\
    get\_factors \\
    hanoi \\
    is\_valid\_parenthesization \\
    knapsack \\
    kth \\
    mergesort \\
    next\_palindrome \\
    next\_permutation \\
    pascal \\
    possible\_change \\
    quicksort \\
    sieve \\
    sqrt \\
    subsequences \\
\end{ttfamily}

\subsection{Appendix E: Data}
\par Note that there may be some values that we collect in the data-collecting code that have been deleted from the data due to irrelevance to the present study’s data analysis process, such as the actual function body, or the NL description given by the model. Nevertheless, the link to the data used is as follows:
\url{https://drive.google.com/file/d/1a5gz7bOJRLWLr5dcCDk6vIMrKEhgxjcK/view?usp=drive_link }

\end{document}